# Enhanced electrical and electrocaloric properties in lead-free $Ba_{0.8}Ca_{0.2}Zr_xTi_{1-x}O_3$ (x = 0 and 0.02) ceramics

Soukaina Merselmiz [a], Zouhair Hanani [a,b], Said Ben Moumen [a], Aleksander Matavž [c], Daoud Mezzane [a,*], Nikola Novak [c], Zahra Abkhar [a], Lahoucine Hajji [a], M'barek Amjoud [a], Yaovi Gagou [d], Khalid Hoummada [e], Dejvid Črešnar [c], Zdravko Kutnjak [c] and Brigita Rožič [c]

[a] IMED-Lab, Cadi Ayyad University, Marrakesh, 40000, Morocco

[b] ICMCB, University of Bordeaux, Pessac, 33600, France

[c] Jozef Stefan Institute, Ljubljana, 1000, Slovenia

[d] LPMC, University of Picardy Jules Verne, Amiens, 80039, France

[e] IM2NP, Aix-Marseille University, UMR 7334, 13397, Marseille, France

*Corresponding author:* daoudmezzane@gmail.com

## Abstract

The effects of 2% Zr introduction in $Ba_{0.8}Ca_{0.2}TiO_3$ (BCT) system on its electrical and the electrocaloric properties was investigated. BCT and $Ba_{0.8}Ca_{0.2}Zr_{0.02}Ti_{0.98}O_3$ (BCZT) ceramics synthesized by solid-state processing. Both samples were crystalized in a pure perovskite phase with group space *P4mm*. After Zr insertion, enhanced dielectric constant was obtained around the Curie temperature ($T_c$) in BCZT ceramic ($\varepsilon_r$=6330 at $T_c$=388 K) compared to BCT ceramic ($\varepsilon_r$=5080 at $T_c$=388.6 K). Moreover, the large-signal piezoelectric coefficient ($d_{33}^*$) was improved from 270 to 310 pm/V in BCT and BCZT ceramics, respectively, under a moderate electric field of 25 kV/cm. The electrocaloric effect was determined via indirect and direct approaches. In the indirect approach, the electrocaloric temperature change (*ΔT*) was calculated via Maxwell relation and the measured ferroelectric polarization *P* (*E, T*) extracted from the *P–E* curves recorded at 24 kV/cm. The maximum of *ΔT* and the electrocaloric responsivity (*ζ*) of BCZT ceramic (*ΔT*=0.68 K and *ζ*=0.283 K.mm/kV at 385 K) were found to be higher than BCT ceramic (*ΔT*=0.37 K and *ζ*=0.154 K.mm/kV at 387 K). In the direct approach, *ΔT* was measured by means of modified high-resolution calorimeter at 14 kV/cm. As the direct method is more sensitive to the latent heat, it provided larger values for smaller applied field, i.e., *ΔT* = 0.474 and 0.668 K for BCT and BCZT ceramics, respectively. A significant *ζ* of 0.477 K.mm/kV was obtained in BCZT at 385 K and 14 kV/cm that matches the values found in lead-based materials. This research suggests that BCZT lead-free ceramics could have a good potential to be used in solid-state refrigeration applications.

## 1. Introduction

The past decade has seen the rapid research and design of new ferroelectric materials with excellent dielectric, ferroelectric, piezoelectric and electrocaloric properties [1–3]. Among these materials, lead-based ceramics like $Pb(Zr,Ti)O_3$ (PZT) own high dielectric and piezoelectric properties [4]. Nevertheless, the usage of lead has been restricted due to its toxicity and damaging to the environment and our health [4–8]. Consequently, it is highly desirable to develop eco-friendly lead-free materials with properties comparable to those of lead-based compounds [4,9–12]. Among reported lead-free materials, barium titanate ($BaTiO_3$, BT) is one of potentially promising Pb-free materials for developing capacitors, sensors, actuators and dielectric cooling devices, due to its enhanced electrical and electrocaloric properties [13–17]. However, as compared to the lead-based perovskite ceramics such as PZT, the pure $BaTiO_3$ shows the relatively low and stable dielectric constant ($\varepsilon_r$) and piezoelectric coefficient ($d_{33}$) [4,6]. To overcome this drawback, site-doping strategy either by replacing the A-site and/or B-site of perovskite structure is an effective way to improve the dielectric and piezoelectric properties of BT material [18].

Doping BT by $Ca^{2+}$ to form $Ba_{1-x}Ca_xTiO_3$ (BCT) solid solutions have been extensively studied and used for several applications in electronic devices owing to their dielectric, ferroelectric and piezoelectric properties. This substitution caused a slight change in the Curie point ($T_c$) with increasing calcium concentration, but strongly lowers the orthorhombic-tetragonal ($T_{O-T}$) transition temperature [19,20], whereas it can improve the temperature stability of the piezoelectric properties for several practical applications [21–24]. Furthermore, the $Zr^{4+}$ introduction in BCT ceramic, to form $Ba_{1-x}Ca_xZr_yTi_{1-y}O_3$ (BCZT) system, expands the perovskite lattice because of the rather large radius difference between $Zr^{4+}$ (0.72 Å) and $Ti^{4+}$ (0.605 Å) [20,25–29]. Consequently, the lattice distortion leads to improve the strain level and enhanced the piezoelectric effect in BCZT ceramic [30,31]. Additionally, enhanced dielectric and ferroelectric properties were reported in BCZT [32,33].

Beside its enhanced electrical properties, BCZT ceramics have potential for application in new generation of solid-state cooling devices based on electrocaloric effect (ECE) [3,34]. Asbani et al. [25] reported the electrocaloric temperature change ($\Delta T$) of 0.18 K near the $T_c$ in $Ba_{0.8}Ca_{0.2}Zr_{0.02}Ti_{0.98}O_3$, under an applied electric field of 7.95 kV/cm, corresponding to an electrocaloric responsivity ($\zeta = \Delta T/\Delta E$) of 0.226 K.mm/kV. Moreover, Singh et al. [35] stated a maximum of $\Delta T$ of 0.38 K and $\zeta$ of 0.253 K.mm/kV near $T_c$ in $Ba_{0.92}Ca_{0.08}Zr_{0.05}Ti_{0.95}O_3$ ceramic under 15 kV/cm. In a related work, Kaddoussi et al. [26] reported a maximum value of $\Delta T$ of 0.2 K at 8 kV/cm ($\zeta=0.25$ K.mm/kV) in $Ba_{0.95}Ca_{0.05}Zr_{0.10}Ti_{0.90}O_3$ ceramic. Nevertheless, the Achilles' heel of the majority of these EC materials is the low electrocaloric responsivity ($\zeta = \Delta T/\Delta E$), which is less than 0.35 K.mm/kV [34,36–38].

In this work, we demonstrate the effect of introducing 2% of $Zr^{4+}$ in $Ba_{0.8}Ca_{0.2}TiO_3$ ceramic on the structural, electrical and electrocaloric properties. These latter were investigated via direct and indirect approaches at moderate electric fields. A significant $\zeta$ of 0.477 K.mm/kV was found in BCZT at 14 kV/cm that competes the values found in lead-based materials. Hence, BCZT ceramics could be a suitable material for applications in solid-state refrigeration technologies.

## 2. Experimental

### 2.1. Elaboration of BCT and BCZT ceramics

$Ba_{0.8}Ca_{0.2}TiO_3$ (BCT) and $Ba_{0.8}Ca_{0.2}Zr_{0.02}Ti_{0.98}O_3$ (BCZT) ceramics were synthesized by conventional solid-state reaction route. The starting materials were barium carbonate ($BaCO_3$, ≥ 99%, VWR Chemicals), calcium carbonate ($CaCO_3$, ≥ 98.5%, VWR Chemicals), zirconium oxide ($ZrO_2$, 99.5%, Merck) and titanium oxide ($TiO_2$, ≥ 99.5%, VWR Chemicals). The precursors for BCT and BCZT materials were weighed and grinded in the desired stoichiometry by using ethanol as a medium in an agate mortar for 2h. Then, BCT and BCZT powders were calcined at 1150 °C/10h and 1250 °C/12h, respectively. The calcined powders were uniaxially pressed into pellets of diameter about 12 mm and thickness about 1 mm, using 5 wt % of polyvinyl alcohol (PVA) as a binder. The samples were first heated up at 800 °C for 2 h to remove the binder, then sintered at 1350 °C for 7h.

### 2.1. Characterizations

Crystalline structure of BCT and BCZT sintered ceramics was determined by the X-ray diffraction (XRD, Panalytical X-Pert Pro) using Cu-K$_\alpha$ radiation ($\lambda \sim 1.5406$ Å). The lattice parameters of the samples were determined and refined by using the FullProf software. The surface morphology of the sintered ceramics was examined by using the Scanning Electron Microscopy (SEM, Tescan VEGA3). The grain size distributions of the samples were determined by using ImageJ software. The density of the sintered ceramics was evaluated by the Archimedes method using deionized water as a medium. The dielectric properties of BCT and BCZT pellets electroded by silver paste were measured by a precision LCR Meter (Hioki, IM 3570) in the frequency range of 20 Hz to 1 MHz. The polarization–electric field and strain–electric field hysteresis loops were measured by using an AixACCT TF 2000 Analyzer with a SIOS Meβtechnik GmbH laser interferometer and a TREK model 609E-6 high-voltage amplifier. The hysteresis loops were acquired by using an excitation sinusoidal signal with frequency of 10 Hz in the temperature range of 303–403 K with a 5 K step upon heating cycle. The direct electrocaloric measurements were measured by high-resolution calorimeter allowing high resolution measurements of the sample-temperature variation due to the *ECE* induced by a change in the applied bias electric field [39,40]. The samples were covered with silver electrodes and the temperature was measured by using a small bead thermistor as described in the reference [41].

## 3. Results and discussion

*3.1. Structural study*

The room-temperature X-ray diffraction patterns of BCT and BCZT sintered ceramics are plotted in Fig. 1a. Both ceramics were crystalized in a pure perovskite phase, without any trace of crystalline impurities. All diffraction peaks can be indexed based on the standard X-ray model of polycrystalline tetragonal $BaTiO_3$ (PDF#79-2264) with the space group *P4mm*. Fig. 1b shows the Gaussian fitted enlarged pattern of the diffraction peaks around $2\theta$= 44 – 46°. The phase analysis of BCT and BCZT sintered ceramics are shown in Figs. 1c and 1d. The refinement performed by the FullProf Suite software revealed that the diffraction data of both ceramics correspond to a tetragonal phase (*P4mm*). Rietveld refinement confirmed the substitution of both $Ca^{2+}$ and $Zr^{4+}$ in A-site and B-site, respectively [25,36,42]. In Table 1, the lattice parameters, atomic positions and space group obtained after refinement are presented. It was observed that the volume of BCT ceramic (62.6963 Å$^3$) was increased after the insertion of $Zr^{4+}$ to form BCZT ceramic (62.9969 Å$^3$). These results are predictable and related to the larger $Zr^{4+}$ ionic radius (0.72 Å) than that of $Ti^{4+}$ (0.605 Å), which increased the lattice parameter of BCZT ceramic [25,33,43,44]. The degree of tetragonality (*c/a*) was decreased after the insertion of $Zr^{4+}$ ion due to the lattice expansion as reported by Chen et al. [44]. It was observed that the (200) and (201) peaks form of BCZT ceramics are stronger compared to that of BCT ceramics. This could be related to preferential crystallographic orientation and/or growth of the crystals in BCT and BCZT ceramics [45].

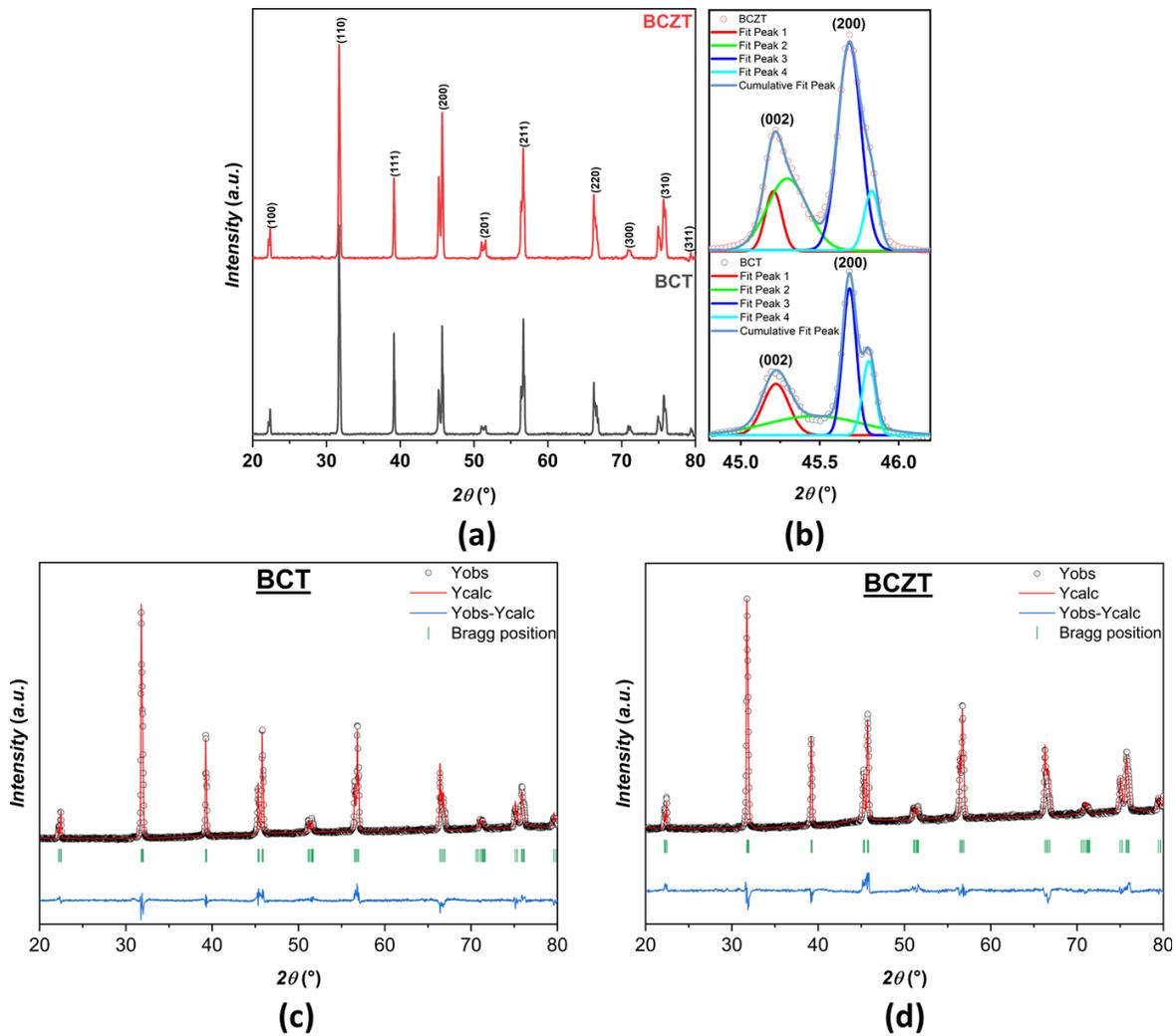

**Fig. 1** (a) XRD patterns, (b) peak deconvolution around $2\theta \approx 45°$ and Rietveld refinement of (c) BCT and (d) BCZT ceramics.

**Table 1** Results of Rietveld refinement parameters of BCT and BCZT ceramics.

| Sample | Structure | Space group | Lattice parameters | Atomic positions (x, y, z) | $\chi^2$ |
|---|---|---|---|---|---|
| **BCT** | Tetragonal | *P4mm* | $a$ = 3.9593 Å<br>$c$ = 3.9995 Å<br>$c/a$ =1.0101<br>$V$ =62.6963 Å$^3$<br>$\alpha = \beta = \gamma = 90$ | Ba/Ca (0,0, -0.0325)<br>Ti (0.5,0.5, 0.4859)<br>O1 (0.5,0, 0.4493)<br>O2 (0.5,0.5, 0.0217) | 1.85 |
| **BCZT** | Tetragonal | *P4mm* | $a$ = 3.9663 Å<br>$c$ = 4.0045 Å<br>$c/a$ =1.0096<br>$V$ = 62.9969 Å$^3$<br>$\alpha = \beta = \gamma = 90$ | Ba/Ca (0,0, -0.0402)<br>Ti/Zr (0.5,0.5,0.4845)<br>O1 (0.5,0,0.40260)<br>O2 (0.5,0.5,0.0204) | 1.74 |

*3.2. Morphological study*

SEM micrographs and grain size distributions of BCT and BCZT ceramics sintered at 1350 °C/7h are displayed in Fig. 2. All samples exhibit compact and inhomogeneous grains with lognormal and gaussian distribution for BCT and BCZT ceramics, respectively. The presence of small and coarse grains (Fig. 2a) with an average grain size of (5.24 ± 3.52) µm (see inset to Fig. 2a) was shown in BCT ceramics. Whereas, BCZT ceramics exhibit homogeneous coarse grains with grain size of (8.42 ± 3.57) µm (Fig. 2b). It should be noted that the substitution of $Ti^{4+}$ by $Zr^{4+}$ affects the microstructure, i.e., stimulates the increase of the grain size in BCZT in comparison to BCT [25,46]. Moreover, better compactness of samples is observed in BCZT with higher bulk density of 5.45 g/cm$^3$ in contrast to 5.20 g/cm$^3$ found in BCT ceramic. This result is consistent with SEM observations.

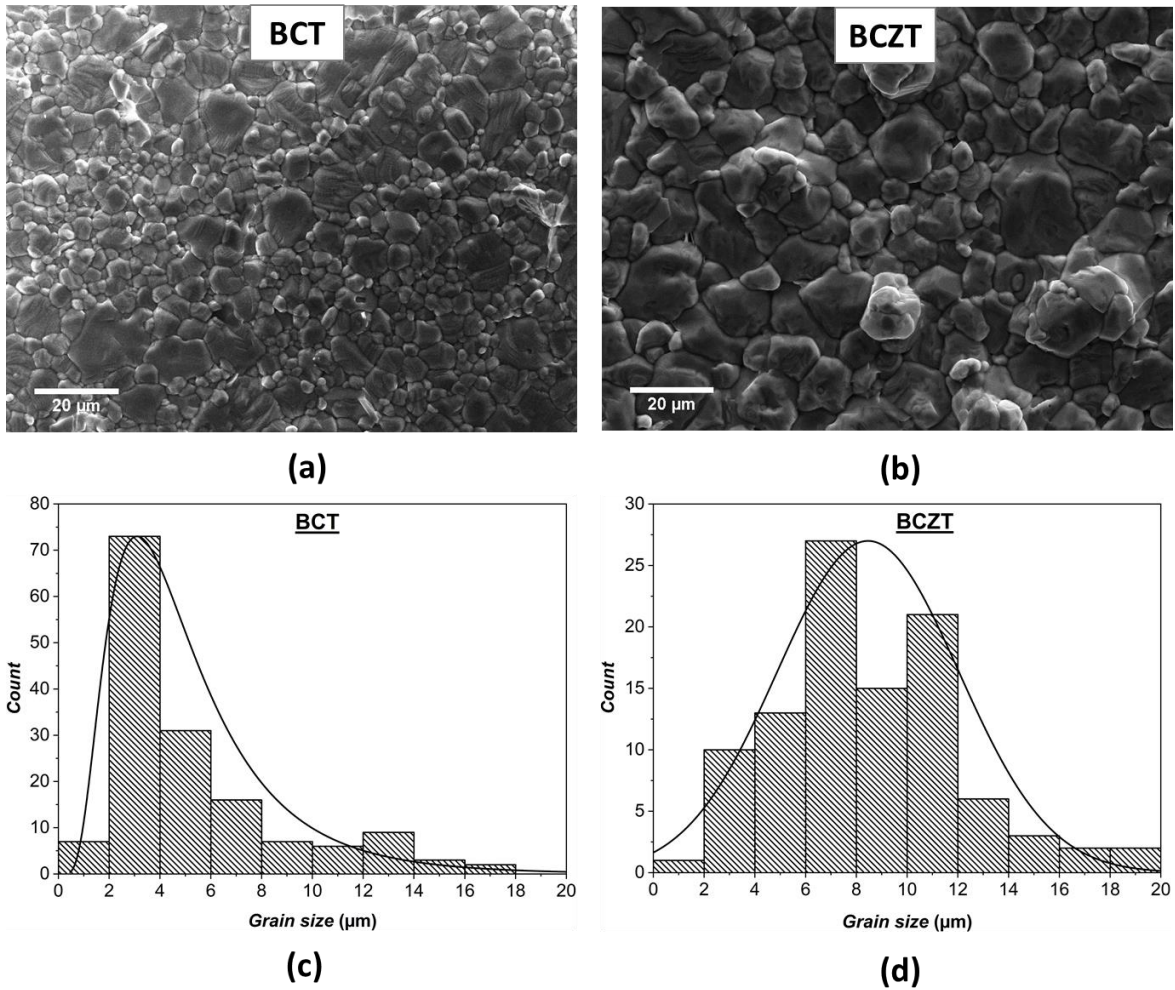

**Fig. 2** (a, b) SEM micrographs and (c, d) grain size distribution of BCT and BCZT ceramics, respectively.

## 3.3. Electrical performances

### 3.3.1 Dielectric properties

Fig. 3 displays the temperature-dependence of the dielectric constant ($\varepsilon_r$) and the dielectric loss ($\tan \delta$) at various frequencies in the BCT and BCZT sintered ceramics. Both samples show only one phase transition around Curie temperature ($T_c$), which is associated to the cubic-tetragonal (C–T) phase transition. At 1 kHz, the maximum of $\varepsilon_r$ was enhanced from 5080 to 6330 in BCT and BCZT ceramics, respectively. However, $\tan \delta$ was slightly increased from 0.029 to 0.043 in BCT and BCZT ceramics, respectively. The increase in $\varepsilon_r$ and $\tan \delta$ could be ascribed to the grain size increasing after Zr insertion, as reported in the literature [33,46–50]. Moreover, $T_c$ was slightly decreased from 388.6 to 388 K in BCT and BCZT ceramics, respectively, due to the incorporation of a low Zr content. The decrease of $T_c$ could be related to weakening of the bonding force between B-site ion and the oxygen ion (B-O bonds) in $ABO_3$ perovskite structure due to the difference between $Ti^{4+}$ and $Zr^{4+}$ radii [36,51]. Besides, the peak temperature of BCZT ceramic does not shift with the frequency change, confirming that both ceramics present a non-relaxor behavior because of the small amount of Zr [33,43]. These results are in good agreement with those reported by Asbani et al. [36].

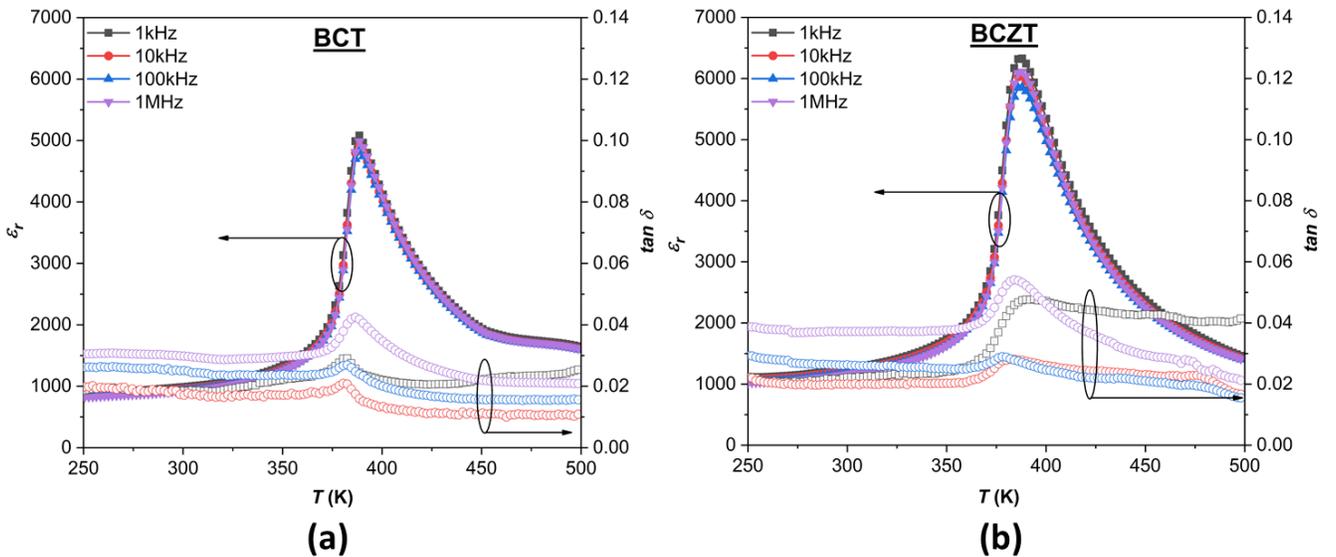

**Fig. 3** Temperature-dependence of the dielectric constant and loss at different frequencies of (a) BCT and (b) BCZT ceramics.

### 3.3.2 Ferroelectric properties

The temperature-dependence of the bipolar *P–E* hysteresis loops measured under 25 kV/cm at 10 Hz are plotted in Fig. 4 (a, b). At room temperature, both BCT and BCZT ceramics show the ferroelectric character of well-saturated hysteresis loops [52], which confirms the ferroelectric nature of the samples. At room temperature, the maximal polarization ($P_{max}$) and the remnant polarization ($P_r$) of BCZT ceramic ($P_{max}$ = 16.01 µC/cm² and $P_r$ = 10.98 µC/cm²) were found to be higher than those of BCT ceramic ($P_{max}$ = 14.60 µC/cm² and $P_r$ =10.72 µC/cm²) due to the substitution of $Ti^{4+}$ by $Zr^{4+}$ as previously reported by Asbani et al. [25]. Meanwhile, the coercive field ($E_c$) of BCZT ceramic ($E_c$= 9.30 kV/cm) was slightly decreased in comparison to BCT ceramic ($E_c$= 10.10 kV/cm). The enhanced ferroelectric properties in BCZT are attributed to the chemical stability and the larger ionic radius of $Zr^{4+}$ compared to $Ti^{4+}$, leading to the expansion of the unit cell of BCT ceramic [33,53]. As increasing the temperature, the *P–E* hysteresis loops gradually become slimmer accompanied by a continuous decrease of $P_r$, $P_{max}$ as well as $E_c$, due to the gradual ferroelectric softening. It was observed that the *P–E* hysteresis loops of both BCT and BCZT ceramics slightly shift with increasing temperature along the field axis, due to the internal bias field possibly originating from some trapped charges on the grain boundaries or defects [54–56].

### *3.3.3 Piezoelectric properties*

Besides the *P–E* hysteresis loops, polarization switching under an electric field in ferroelectric materials leads to strain hysteresis loops, i.e., (*S–E*) hysteresis loops. Fig. 4 (c, d) display the *S–E* curves obtained at 25 kV/cm from 303 to 403 K. At room temperature, typical butterfly-shaped *S–E* curves were observed in both ceramics, alongside with a large negative strain related to the domain switching [57,58]. Such a butterfly loop behavior is due to the normal converse piezoelectric effect of the lattice along with the switching and movement of domain walls by the electric field [52,57,59]. It is interesting to note that the butterfly-shaped strain is large and have an asymmetric nature for positive and negative electric field cycle, which may be related to the internal bias field arising from defects [60]. The maximal average strain ($S_{max}$) at the maximal applied electric field $E_{max}$ for BCT and BCZT ceramics are 0.056 and 0.054 %, respectively. As increasing the temperature, the negative strain decreases gradually, causing the transformation from typical butterfly-shaped to sprout-shaped strain curves, due to the gradual disappearance of ferroelectricity [61–63].

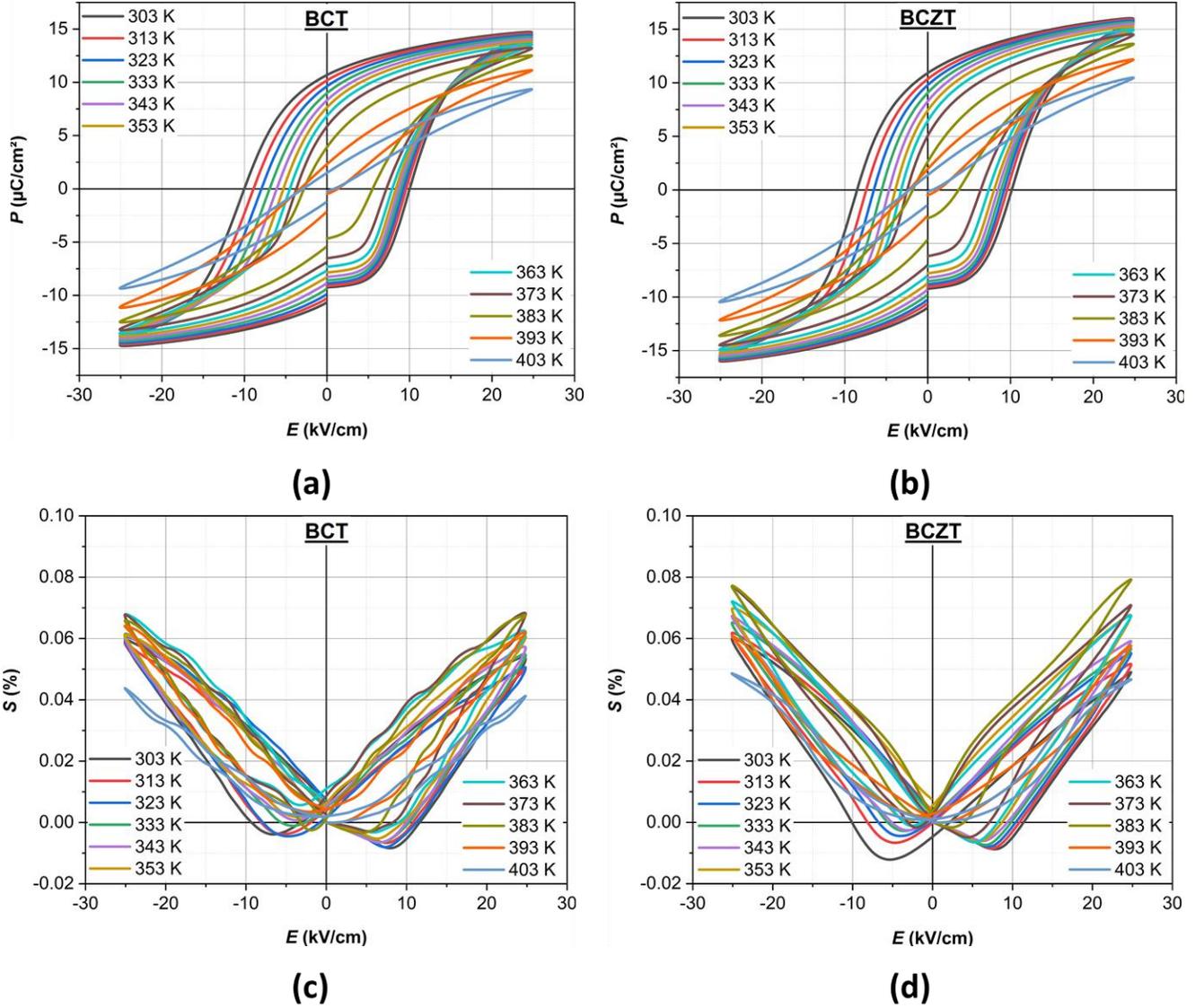

**Fig. 4** (a, b) *P–E* and (c, d) *S–E* hysteresis curves acquired at different temperatures in BCT and BCZT ceramics.

The large signal piezoelectric coefficient also denoted as normalized strain ($d_{33}^*$) is defined by $d_{33}^* = S_{max}/E_{max}$, where $S_{max}$ is the maximum strain at the maximum electric field $E_{max}$. $d_{33,ave}^*$ is the average value of the $d_{33}^*$ calculated from $S_{max}$ measured at positive and negative $E_{max}$. The temperature profiles of $d_{33,ave}^*$ for both samples are shown in Fig. 5. At room temperature, $d_{33,ave}^*$ of BCT ceramic (226.9 pm/V) is higher than BCZT ceramic (218.6 pm/V). Nevertheless, as increasing the temperature, $d_{33,ave}^*$ was increased gradually toward the FE-PE phase transition to reach a maximum of 270 and 310 pm/V in BCT and BCZT ceramics, respectively. Above the peak temperature, the $d_{33,ave}^*$ decreases sharply in both ceramics due to the onset of the paraelectric phase. The enhanced $d_{33,ave}^*$ in BCZT ceramic could be related to lattice distortion after Zr insertion, which lead to enhance the strain level and then the piezoelectric effect [30,31].

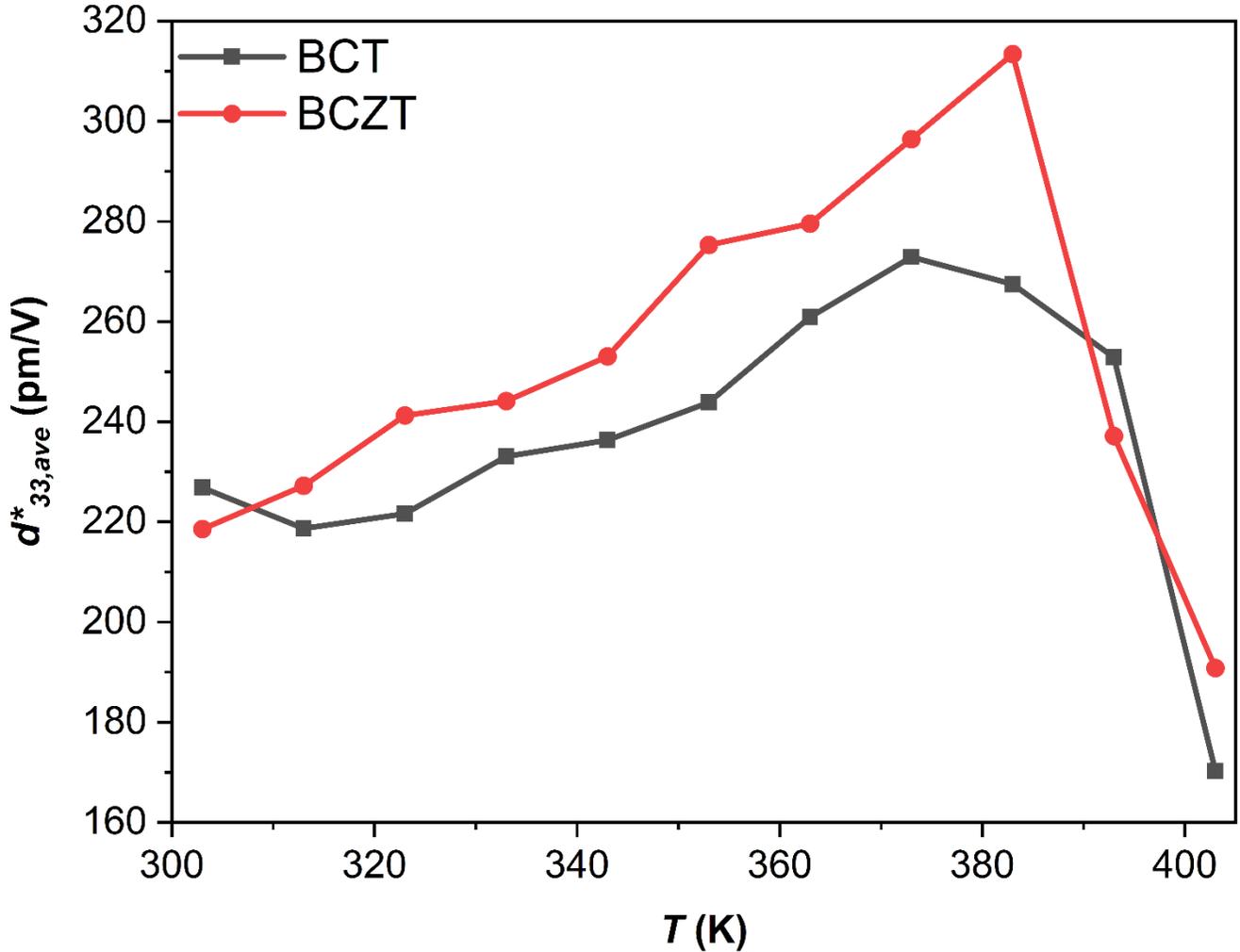

**Fig. 5** Thermal evolution of $d^*_{33,ave}$ in BCT and BCZT ceramics.

At 30 kV/cm, Chaiyo et al. [64] reported that $Ba_{0.75}Ca_{0.25}TiO_3$ and $Ba_{0.80}Ca_{0.20}Zr_{0.05}Ti_{0.95}O_3$ ceramics exhibit $d^*_{33}$ of 173 pm/V and 284 pm/V, respectively. These values are comparable with those obtained in BCT and BCZT samples. Moreover, Pisitpipathsin et al. [20] stated that $Ba_{0.91}Ca_{0.09}Zr_{0.04}Ti_{0.96}O_3$ ceramic displays a maximum of $d^*_{33}$=513 pm/V at $T_c$=123 °C under 60 kV/cm. The differences in these results could be attributed to the chemical compositions, the applied electric field, and the measurement conditions.

*3.4. Electrocaloric properties*

To evaluate the EC effect in lead-free BCT and BCZT ceramics for environmentally friendly solid-state cooling devices, the measurements of the electrocaloric effect were performed by employing two methods, (i) the indirect experimental method following the Maxwell relation and measured $P(E, T)$ and (ii) the direct method by means of high-resolution calorimeter. Moreover, the results of both methods were compared and discussed.

### 3.4.1. Indirect method

The indirect method basing on the Maxwell approach for EC effect evaluation is based on the measured ferroelectric order parameter $P(E, T)$ determined from $P$–$E$ hysteresis curves (see Figs. 4a and 4b). After, a seven-order polynomial fitting of only the upper branches of these $P$–$E$ hysteresis curves at every fixed applied electric fields, the variation of the polarization versus temperature was deduced (plotted in Figs. 6a and 6b). For both ceramics, the polarization decreases slowly toward $T_c$, then drops gradually. The reversible electrocaloric temperature change ($\Delta T$) was calculated via an indirect method following the Maxwell equation,

$$\Delta T = -\int_{E_1}^{E_2} \frac{T}{\rho C_p} \left(\frac{\partial P}{\partial T}\right)_E dE, \qquad (1)$$

Here $\rho$ and $C_p$ are the mass density and the specific heat of the material, respectively.

The temperature-dependent of $\Delta T$ obtained at different electric fields for BCT and BCZT ceramics are shown in Figs. 6c and 6d, respectively. Each curve that corresponds to a fixed applied electric field exhibits a maximum around the ferroelectric-paraelectric (FE-PE) phase transition. The highest values of $\Delta T$ were found to be of 0.37 K and 0.68 K under 24 kV/cm near the FE-PE transition of BCT and BCZT ceramics, respectively. It should be noted that further increase in $\Delta E$ could even more enhance the $\Delta T$ value of both BCT and BCZT ceramics, since no saturation in $\Delta T$ was reached yet in both materials. Consequently, the electrocaloric responsivity ($\zeta = \Delta T/\Delta E$) is in such case more suitable to evaluate the strength of electrocaloric response. At 24 kV/cm, the electrocaloric responsivity ($\zeta$) of 0.154 and 0.283 K.mm/kV were estimated in BCT and BCZT ceramics, respectively. The enhanced $\Delta T$ in BCZT ceramic could be ascribed to the improved ferroelectric properties after Zr incorporation.

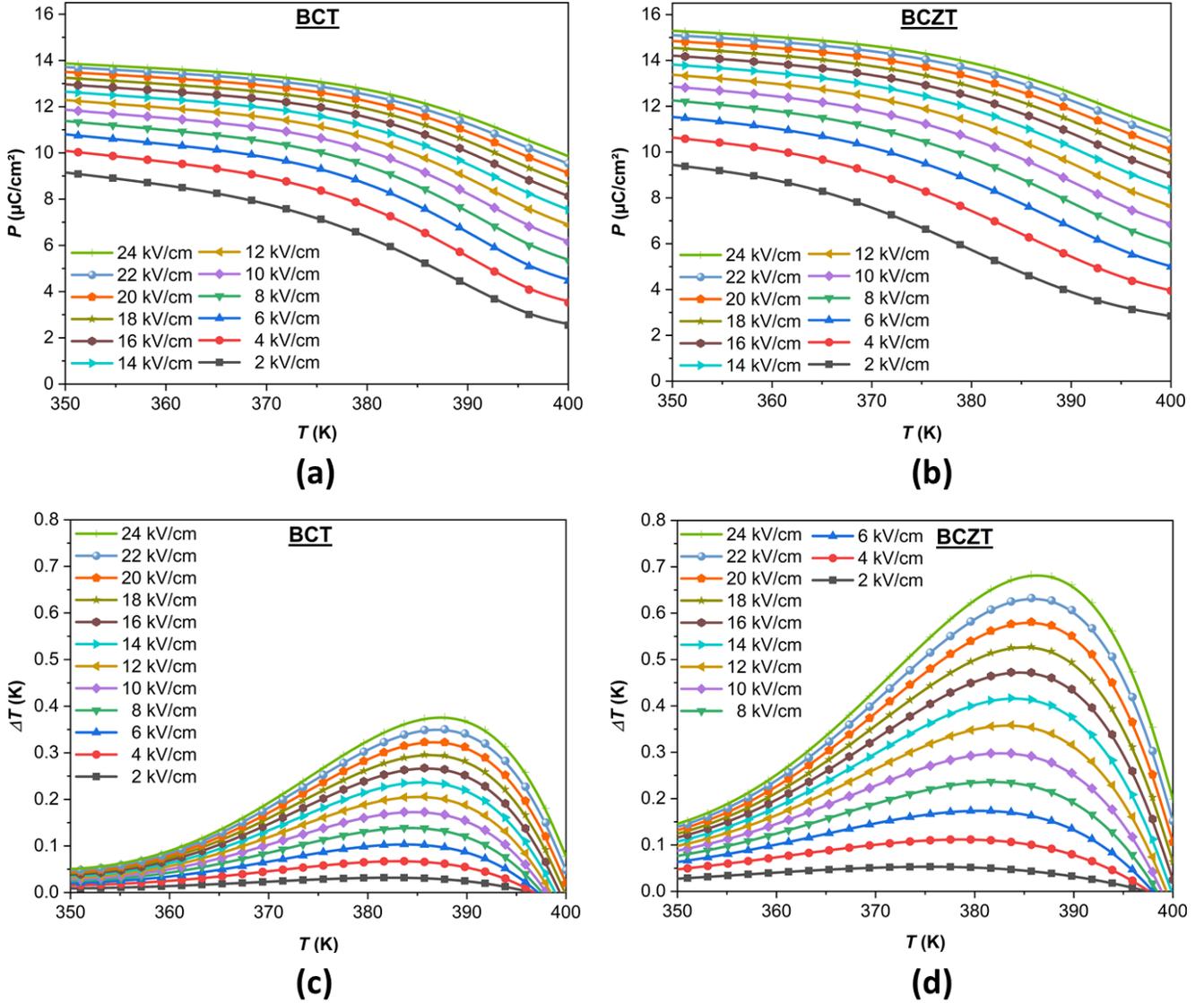

**Fig. 6** Temperature-dependence of (a, b) polarization and (c, d) electrocaloric temperature change ($\Delta T$) determined at different applied electric fields in BCT and BCZT ceramics.

To situate our finding to literature, Table 2 compares the electrocaloric properties ($\Delta T$ and $\zeta$) of BCT and BCZT ceramics with previously published results obtained in lead-free ferroelectric ceramics. The $\Delta T$ values of BCT and BCZT ceramics presented here are among the highest so far reported in literature. In particular, at 8 kV/cm and at the same chemical composition, the $\Delta T$ of BCT ceramic ($\Delta T$= 0.14 K) and BCZT ceramic ($\Delta T$= 0.24 K) are larger than those obtained by Asbani et al. [25,36]. Meanwhile, Kaddoussi et al. [26] studied the EC effect in $Ba_{0.85}Ca_{0.15}Zr_{0.10}Ti_{0.90}O_3$ and found that $\Delta T$ reached 0.152 K at 8 kV/cm, which is lower than our results. Moreover, the highest $\Delta T$ of 0.67 K obtained indirectly in BCZT at 24 kV/cm is higher to that found by Ben Abdessalem et al. [65] ($\Delta T$ = 0.565 K) at 30 kV/cm. The differences in these results could be attributed not only to the elaboration conditions, but also to the chemical composition, the number of coexisting phases and the applied electric field.

*3.4.2. Direct measurements*

The direct electrocaloric response ($\Delta T$) as a function of temperature was examined by using a high-resolution calorimeter at applied electric field of 6, 8, 10, 14 kV/cm, as shown in Figs. 7a and 7b. At 14 kV/cm, $\Delta T$ of 0.474 and 0.668 K was obtained in BCT and BCZT ceramics at $T_c$, respectively. The corresponding $\zeta$ values of 0.338 and 0.477 K.mm/kV at 14 kV/cm were found in BCT and BCZT ceramics, respectively. Hence, introducing 2% of $Zr^{4+}$ into BCT lattice, enhanced the electrocaloric properties of BCT ceramics [25]. Unfortunately, voltage limitations in the direct EC effect measurement set-up prevent us to apply higher electric fields above 14 kV/cm.

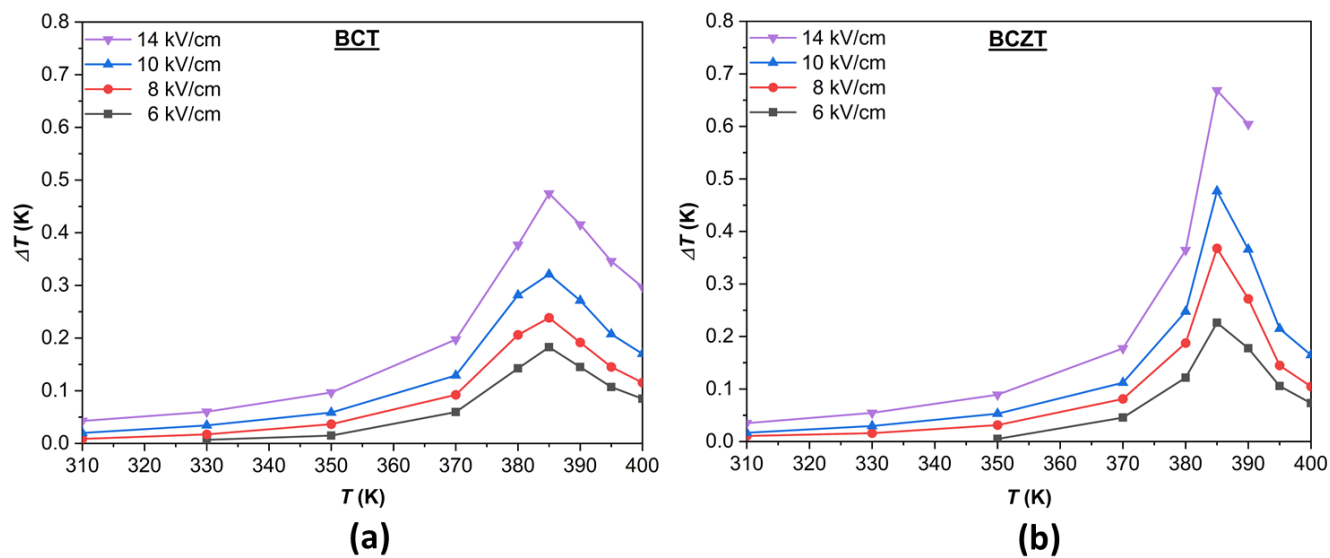

**Fig. 7** Temperature-dependence of $\Delta T$ obtained by using the direct measurement method for (a) BCT and (b) BCZT ceramics at several electric fields.

Fig. 8 presents the comparison between the EC effect ($\Delta T$ and $\zeta$) results obtained by both indirect and direct methods at the same electric field of 14 kV/cm. Both methods confirm the tendency of the electrocaloric effect to exhibit a peak at the phase transition. However, as anticipated, the indirect method shows typically smeared and slightly suppressed anomalies around the phase transition temperature. In contrast, the direct method shows much sharper and higher peaks near the phase transition temperature. While the direct method is highly sensitive to the latent heat absorbed/released at the sharp first order transitions and can easily follow sharpness of the second order transitions [17], the indirect method tends to smear significantly the EC effect transition anomalies due to its intricate fitting-procedure steps. Our results thus demonstrate that in the case of sharp phase transitions, the EC effect results obtained by indirect method should be taken by caution.

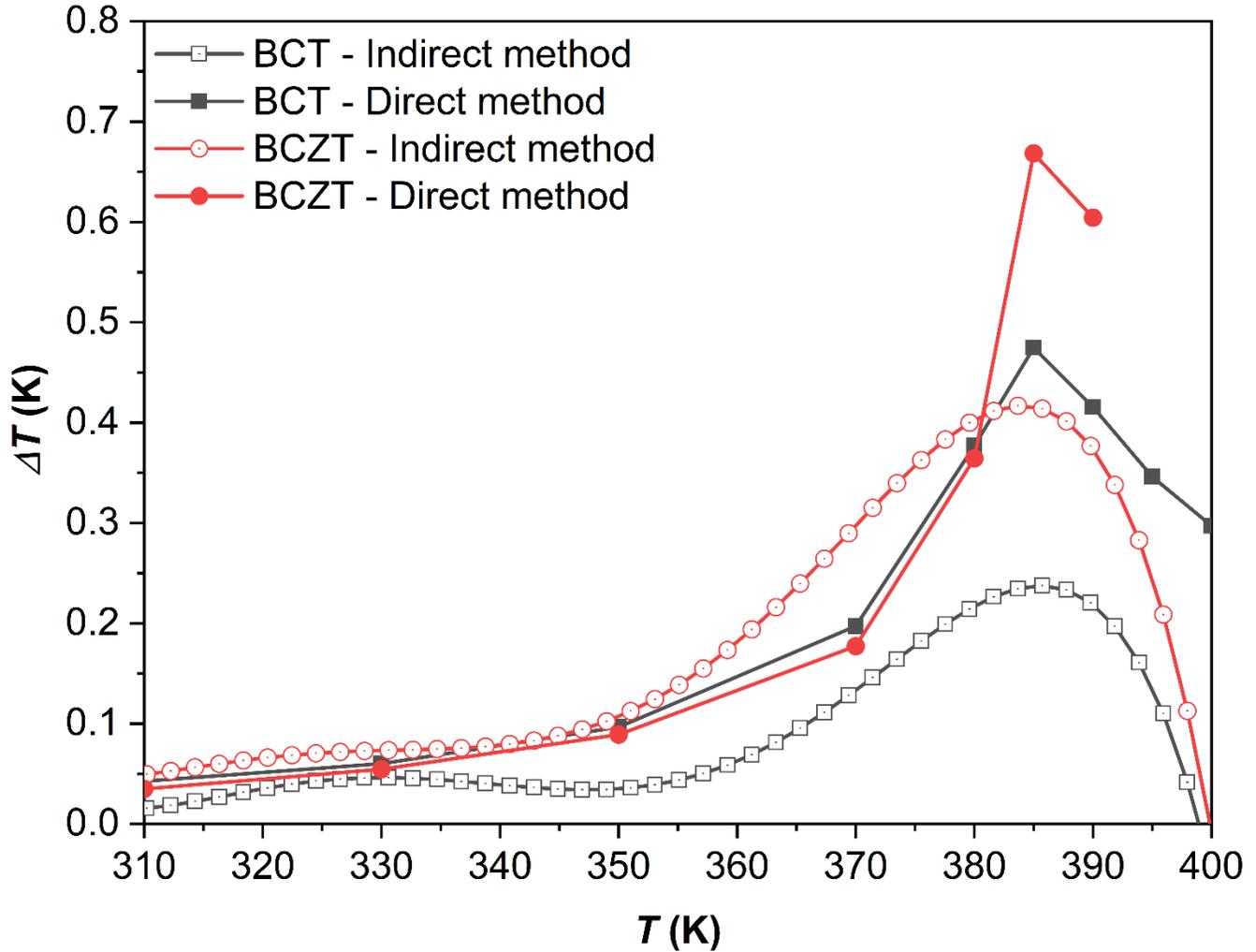

**Fig. 8** Comparison of $\Delta T$ obtained by indirect method (open symbols) and by direct method (solid symbols) at 14 kV/cm for BCT and BCZT ceramics.

Table 2 compares the electrocaloric properties of lead-free BCT and BCZT ceramics with previously published results. At 8 kV/cm, the $\Delta T$ values obtained by direct method are 0.23 and 0.36 K for BCT and BCZT ceramics, respectively, which presents the highest values compared to those reported in literature. At the same electric field, Kaddoussi et al. [26] measured the ECE directly for $Ba_{0.95}Ca_{0.05}Zr_{0.1}Ti_{0.9}O_3$ ceramics and found a maximum of $\Delta T$ of 0.25 K around 360 K with $\zeta$ of 0.31 K.mm/kV. Besides, Hanani et al. [3] reported *EC* responses under 17 kV/cm ($\Delta T$ = 0.492 K and $\zeta$ = 0.289 K.mm/kV at 360 K) in rod-like $Ba_{0.85}Ca_{0.15}Zr_{0.10}Ti_{0.90}O_3$ ceramic elaborated by surfactant-assisted solvothermal route. It was reported by Sanlialp et al. [66] that BZT-32BCT ceramics exhibits at 20 kV/cm, a maximum of $\Delta T$ and $\zeta$ values of 0.33 K and 0.165 K.mm/kV, respectively, around 337 K. In our case, by applying an external electric field of only 14 kV/cm on BCZT ceramics, larger $\Delta T$ equal to 0.668 K and higher $\zeta$ of 0.477 K.mm/kV around 385 K were obtained. These values are one of the highest values reported for lead-free ferroelectric materials as seen in Table 2. Hereafter, for solid-state cooling devices, the highest temperature change $\Delta T$ under a lowest electric field are needed, consequently BCT and BCZT could be a potential candidate for next generation solid-state cooling devices based on the ECE.

**Table 2** Comparison of the electrocaloric properties obtained by indirect and direct methods for BCT and BCZT ceramics with other lead-free ceramics reported in literature.

| Ceramic | T (K) | $\Delta T_{max}$ (K) | $\Delta E_{max}$ (kV/cm) | $\zeta$ (K.mm/kV) | Measurement Method | Ref. |
|---|---|---|---|---|---|---|
| BCT | 385 | 0.14 | 8 | 0.175 | Indirect | This work |
| BCZT | 381 | 0.24 | 8 | 0.30 | Indirect | This work |
| BCT | 387 | 0.37 | 24 | 0.154 | Indirect | This work |
| BCZT | 385 | 0.68 | 24 | 0.283 | Indirect | This work |
| BCT | 385 | 0.23 | 8 | 0.287 | Direct | This work |
| BCZT | 385 | 0.36 | 8 | 0.45 | Direct | This work |
| BCT | 385 | 0.474 | 14 | 0.338 | Direct | This work |
| BCZT | 385 | 0.668 | 14 | 0.477 | Direct | This work |
| $Ba_{0.85}Ca_{0.15}Zr_{0.10}Ti_{0.90}O_3$ | 363 | 0.115 | 6.65 | 0.164 | Indirect | [34] |
| $Ba_{0.80}Ca_{0.20}TiO_3$ | 398 | 0.12 | 7.95 | 0.15 | Indirect | [36] |
| $Ba_{0.80}Ca_{0.20}Zr_{0.02}Ti_{0.98}O_3$ | 403 | 0.18 | 7.95 | 0.226 | Indirect | [25] |
| $Ba_{0.80}Ca_{0.20}Zr_{0.04}Ti_{0.96}O_3$ | 386 | 0.27 | 7.95 | 0.34 | Indirect | [36] |
| $Ba_{0.95}Ca_{0.05}Zr_{0.10}Ti_{0.90}O_3$ | 368 | 0.205 | 8 | 0.256 | Indirect | [26] |
| $Ba_{0.85}Ca_{0.15}Zr_{0.10}Ti_{0.90}O_3$ | 373 | 0.152 | 8 | 0.19 | Indirect | [26] |
| $Ba_{0.95}Ca_{0.05}Zr_{0.1}Ti_{0.9}O_3$ | 360 | 0.25 | 8 | 0.31 | Direct | [26] |
| $Ba_{0.92}Ca_{0.08}Zr_{0.05}Ti_{0.95}O_3$ | 410 | 0.38 | 15 | 0.253 | Indirect | [35] |
| $Ba_{0.85}Ca_{0.15}Zr_{0.10}Ti_{0.90}O_3$ | 360 | 0.492 | 17 | 0.289 | Indirect | [3] |
| BZT-30BCT | 333 | 0.30 | 20 | 0.15 | Indirect | [67] |
| BZT-35BCT | 298 | 0.33 | 20 | 0.165 | Direct | [68] |
| BZT–32BCT | 337 | 0.33 | 20 | 0.165 | Direct | [66] |
| $Ba_{0.9}Ca_{0.1}Zr_{0.05}Ti_{0.95}O_3$ | 392 | 0.465 | 25 | 0.186 | Indirect | [65] |

| | | | | | |
|---|---|---|---|---|---|
| $Ba_{0.9}Ca_{0.1}Zr_{0.05}Ti_{0.95}O_3$ | 392 | 0.565 | 30 | 0.188 | Indirect | [65] |

## 4. Conclusions

Zirconium insertion in BCT ceramics enhanced the dielectric, ferroelectric and piezoelectric properties as well as the electrocaloric properties. The latter was investigated by two methods; (i) the indirect calculations from measured $P$ ($E$, $T$) and (ii) the direct electrocaloric temperature change ($\Delta T$) measurements. The results of both methods are compared and it was found that in BCT and BCZT ceramics, the $\Delta T$ values obtained by the direct method are typically larger in comparison to indirect method, which is due to the smearing tendency of latter method. Therefore, the electrocaloric responsivity ($\zeta$) deduced from direct method was found to be 0.338 and 0.477 K.mm/kV for BCT and BCZT, respectively, which are larger values compared to other results obtained in lead-free materials reported in literature and comparable to best lead-based materials. These results make BCZT a potential candidate for integration into future solid-state electrocaloric cooling applications.

**Acknowledgements**

The authors gratefully acknowledge the generous financial support of CNRST Priority Program PPR 15/2015, the Slovenian research agency grants P1-0125, J1-9147, and the European Union's Horizon 2020 Research and Innovation Program under the Marie Skłodowska-Curie Grant Agreement No. 778072.